# Comment on "Pentadiamond: A Hard Carbon Allotrope of a Pentagonal Network of sp$^2$ and sp$^3$ atoms"

V.V. Brazhkin[1*], M.V. Kondrin[1], A.G. Kvashnin[2], E. Mazhnik[2], and A.R. Oganov[2*]

[1] *Institute for High Pressure Physics, 108840 Troitsk, Moscow, Russia*

[2] *Skolkovo Institute of Science and Technology, 121205 Moscow, Russia*

In a recent Letter [1] Y. Fujii et al. proposed a new carbon allotrope and claimed its extremely high Young's and shear moduli of 1691 and 1113 GPa, respectively, surpassing those of diamond by 1.5 and 2 times! The authors also claimed unique negative averaged Poisson's ratio of -0.24, bulk modulus of 381 GPa and record value of the sound speed 28.7 km/s for this carbon modification.

Here we show by two independent computations that these values were obtained as a result of errors. The new hypothetical allotrope has in fact quite moderate elastic moduli, unremarkable for a carbon structure with similar density: bulk modulus of 250 ± 1 GPa; shear modulus of 170 ±2 GPa; Young's modulus of 417 ± 4 GPa, positive Poisson's ratio of 0.22. Even without detailed computations it is obvious that the results of [1] are erroneous: it is impossible to construct the structure made of sp$^2$ and sp$^3$ carbon atoms with bulk or shear modulus exceeding those of diamond or its polytypes [2].

Carbon can exhibit different types of hybridization, which leads to the formation of a large number of different crystalline and amorphous modifications. Finding new hypothetical structures is interesting because of their potentially interesting mechanical, optical and electronic properties. Some of these modifications may have elastic moduli close to those of diamond [2,3], some may exceed the density of diamond by a couple percent [3]. The majority of 3D-linked crystal and amorphous structures are based on sp$^3$ and sp$^2$ carbon atoms have density of about 2 - 3.5 g/cm$^3$, while the elastic moduli of various carbon modifications are in the range from 100 to 450 ГПа [2,3]. It has been computationally proven that diamond is the hardest possible phase of carbon [4] and, more recently, it was proven that it is the hardest possible material among all elements and compounds [5].

A new hypothetical modification of carbon, named pentadiamond (because of the presence of pentagonal rings in its structure) has been proposed and theoretically investigated in [1]. Its structure is cubic (space group Fm3m), contains 22 atoms (10 and 12 of which are sp$^3$ and sp$^2$, respectively) in the primitive cell, and has lattice parameter 9.195 Å. The new hypothetical allotrope is predicted to be extremely unstable (275 meV/atom higher than diamond in energy), which makes its synthesis unlikely – but nevertheless, this new modification can be interesting if it possesses unique properties. Calculations of [1] yielded the bulk modulus *B* equal to 381 GPa, shear modulus *G* of 1113 GPa, and Young's modulus *Y* of 1691 GPa. These values look unbelievable. It is known that the bulk modulus is determined by mean electron density and bond energy [2]. For carbon materials the maximum bulk modulus can be evaluated simply as the bulk modulus of diamond times the ratio of densities of the hypothetical allotrope and diamond. The bulk modulus of pentadiamond (density 2.26 g/cm$^3$) cannot be significantly greater than 290 GPa. The shear modulus is determined not only by the electron density, but also by degree of its localization [2]. The shear modulus of 1113 GPa implies that the degree of electron localization on bonds and angular rigidity in pentadiamond should be 3-4 greater than in diamond, which seems extremely implausible. The predicted Poisson's ratio $\sigma = -0.24$ also appears unlikely. For the known and studied hypothetical 3D allotropes of carbon the Poisson's ratio is always in the range 0.05 – 0.25 [2]. Materials with negative Poisson's ratio (called auxetics), such as α-cristobalite, are very few – and at least in some directions have positive Poisson's ratio [6], while in pentadiamond it is strongly negative in all directions, between -0.2 and -0.28. Such behavior has not been reported on any other matetial. The authors of [1] do not discuss the hardness. Using the formula [7] and values $B$=381 GPa and $G$=1113 GPa, we obtain the Vickers hardness $H$ =422 GPa, which exceeds the hardness of diamond by 4 times. However, recently we found [8] that the formula [7] unexpectedly fails for auxetics. Using the improved formula [8], we obtain the hardness of 210 GPa, i.e. still twice the value for diamond.

Driven by curiosity and doubt, we have calculated the elastic properties of pentadiamond by three ways – using *ab initio* codes QuantumESPRESSO [9] and VASP [10-12] and using our new machine learning model for fast predicting elastic tensors. All three calculations were performed independently. In both *ab initio* calculations we used the generalized gradient approximation for exchange-correlation [13] and the projector augmented-wave method [14,15] for describing core-valence interactions. The plane wave kinetic energy cutoffs were 40 Ry (in QuantumESPRESSO) and 550 eV (in VASP). Integration over Brillouin zone was done using 4×4×4 Monkhorst-Pack grid (in QuantumESPRESSO) and Γ-centered 8x8x8 grid (in VASP). After relaxing the crystal structure, we applied strains of +1% and -1% and extracted the elastic constants from stress-strain relations. Machine learning calculations are extremely fast and establish statistically the likeliest result, based on knowledge of the elastic constants calculated before. Vickers hardness of pentadiamond was estimated according to model [8].

*Table 1 summarizes our results in comparison with [1]. Three independent estimates obtained here give a consistent picture. Ab initio calculations yield the bulk modulus of B = 250 ± 1 GPa, shear modulus G = 170 ±2 GPa, Young's modulus Y = 417 ± 4 GPa, Poisson's ratio σ = 0.22 (see*

Table *1*). Based on these numbers, we estimate the Vickers hardness $H$ = 20±1 GPa. We have also calculated the bulk modulus from the equation of state (computed in QuantumESPRESSO – see Fig. 1), which gave us $B$ = 257 GPa, $B'$ = 3.26, perfectly consistent with the above numbers. Calculations performed at local density approximation level gave similar results ($B$ = 268 GPa, $G$ = 171 GPa, $Y$ = 422 GPa, $\sigma$ = 0.24, $H$ = 19 GPa). All of these properties are unremarkable and in stark contrast with those published in [1].

Table 1. Calculated elastic properties of pentadiamond in comparison with Y. Fujii et al. [1].

| Property | Y. Fujii et al. [1] | This work (Quantum ESPRESSO) | This work (VASP) | This work (machine learning) |
|---|---|---|---|---|
| a, Å | 9.195 | 9.184 | 9.191 | 9.195 |
| E-E(diam), meV/atom | 275 | 263 | 267 | - |
| $C_{11}$, GPa | 1715.3 | 539 | 537 | 409 |
| $C_{12}$, GPa | -283.5 | 105 | 106 | 118 |
| $C_{44}$, GPa | 1187.5 | 141 | 143 | 200 |
| B, GPa | 381 | 250 | 249 | 215 |

| Property | Y. Fujii et al. [1] | This work (Quantum ESPRESSO) | This work (VASP) | This work (machine learning) |
|---|---|---|---|---|
| G, GPa | 1113 | 172 | 169 | 176 |
| Y, GPa | 1691 | 420 | 413 | 415 |
| σ | -0.241 | 0.22 | 0.22 | 0.18 |
| $H_v$, GPa | 210 | 20 | 20 | 26 |

Clearly, elastic constants of pentadiamond reported in [1] are erroneous. At the same time, calculated phonon dispersion curves and electronic band structure reported in [1] are possibly correct. Moreover, the slope of the acoustic phonon branches (given in Fig. 2 in [1]) corresponds to moderate sound velocities 10-11 km/s, much lower than those in diamond, in contrast to the claimed value 28.7 km/s.

We have many times suggested to the authors and editors of journals to be extremely careful with claims of hypothetical and real materials with hardness or elastic moduli exceeding those of diamond [2,16]. There is no physical or chemical basis to support claims of the elastic moduli or hardness of carbon materials exceeding those of diamond by several times. If, however, a paper with such claims is sent to referees, it is better if a referee checks validity of such results whenever possible.

Acknowledgments:

V.V.B. and M.V.K. are grateful to RSF (19-12-00111) for financial support. A.R.O. thanks Russian Ministry of Science and Higher Education (grant 2711.2020.2 to leading scientific schools) for financial support.

[*] Corresponding authors: brazhkin@hppi.troitsk.ru, a.oganov@skoltech.ru

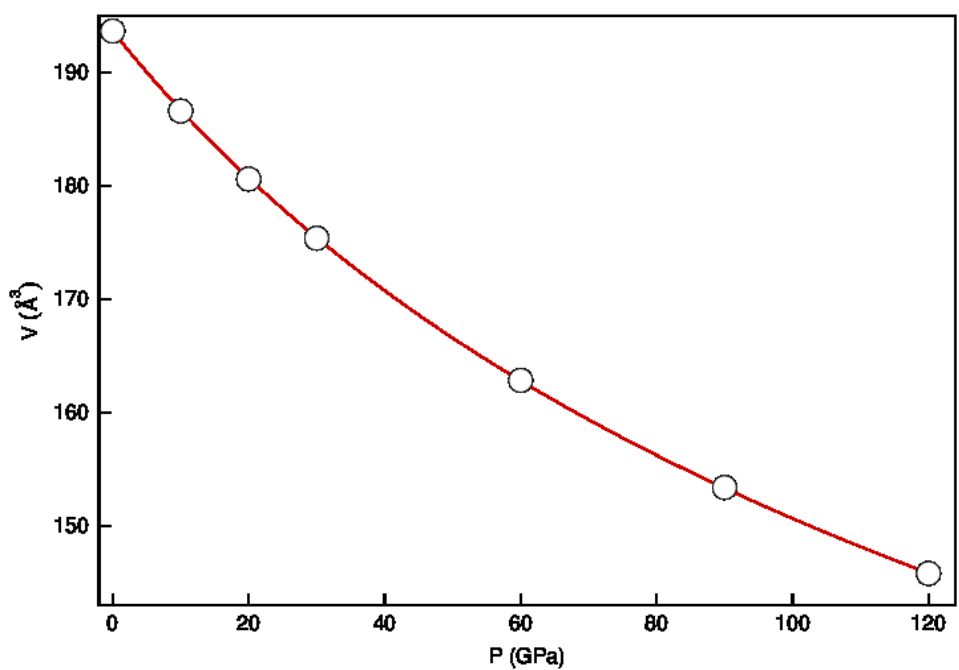

Fig. 1. The calculated compression curve of pentadiamond (volume per unit cell vs pressure)